\documentclass[prl,twocolumn,notitlepage,longbibliography]{revtex4-1}
\usepackage{amsmath}
\usepackage{amssymb}

\usepackage[unicode=true,colorlinks=true,citecolor=blue,urlcolor=blue]{hyperref}

\usepackage{bm}
\usepackage{color}
\usepackage{epsfig}

\usepackage[normalem]{ulem}

\renewcommand {\phi}{{\varphi}}
\newcommand {\rmi}{{\rm i}}

\newcommand {\e}{{\rm e}}

\begin{document}
\title{Topological spin phases of trapped Rydberg excitons in Cu$_2$O}
\author{A.N. Poddubny and M.M. Glazov}
\affiliation{Ioffe Institute, St. Petersburg 194021, Russia}
\email{poddubny@coherent.ioffe.ru}
\email{glazov@coherent.ioffe.ru}
\date{\today}
\begin{abstract}
We study theoretically  Rydberg excitons in one-dimensional chains of traps in Cu$_2$O coupled via the van der Waals interaction. The triplet of optically active $p$-shell states acts as an effective spin{-}$1$ and the interactions between the excitons are strongly spin-dependent.
We predict that the system {has the} topological Haldane phase with {the} diluted antiferromagnetic order, long-range string correlations, and finite  excitation gap. We also analyze the effect of the trap geometry and interactions anisotropy on the Rydberg exciton spin states and demonstrate that 
a rich spin phase diagram can be realized showing high tunability of the Rydberg exciton platform.
\end{abstract}
\maketitle 

{\it Introduction.}  Rydberg states of matter attract high interest nowadays. Strongly excited atoms are macroscopic quantum objects highly susceptible to external fields that serve for bench-top studies of quantum effects at a large scale~\cite{gallagher2005rydberg}. Enhanced polarizability of Rydberg atoms results in efficient interactions making them a platform of choice for quantum simulations~\cite{Saffman2010}.

Exciton, the Coulomb interaction correlated electron-hole pair, emerges in semiconductors when an electron is promoted optically from the filled valence to an empty conduction band. It is a direct analogue of the hydrogen-like atom~\cite{excitons:RS}. Perfectness of the natural cuprous oxide crystals has not only made it possible to discover large-radius excitons in semiconductors~\cite{gross:exciton:eng} but  has lead also to a recent breakthrough: demonstration of stable highly-excited excitonic Rydberg states with principal quantum number $n$ up to 25~\cite{Kazimierczuk2014}. By contrast to Rydberg atoms excitons exist in the crystalline environment, which manifests itself not only in quantitative differences of the binding energy and exciton radii, but also in different selection rules for optical transitions~\cite{PhysRevB.93.075206,Glazov2017,Semina2018} making $p$-shell excitons with the orbital angular momentum $1$ active in single-photon processes, unusual fine structure of the energy spectrum~\cite{Glazov2015}, as well as broken symmetries~\cite{Assmann2016,PhysRevLett.118.046401}. All this provides flexibility to control the excitonic states by light. 

Even more rich consequences of the non-zero orbital angular momentum are expected in exciton-exciton interactions in cuprous oxide~\cite{Walther2018}, which, just like for atoms, are of paramount importance for Rydberg excitons physics~\cite{Kazimierczuk2014}. By now, the interactions are scarcely studied, however, according to the  recent theoretical predictions~\cite{Walther2018},   the coupling between the excitonic states crucially depends on the mutual orientations of their angular momenta.

Similarly to atomic physics, the interaction effects are expected to be strongest in ensembles of localized Rydberg excitons.  Here we demonstrate that already in a one-dimensional {chain} of trapped Rydberg excitons in Cu$_2$O the ground state corresponds to a topologically non-trivial spin order --- the Haldane phase~\cite{Haldane1983} --- with diluted antiferromagnetic order and a gapped spectrum of elementary excitations. Such a topological phase is inherent to integer spin, in particular spin-$1$, chains with antiferromagnetic coupling.
The hallmarks of the Haldane phase are the  non-trivial edge states behaving akin spin-$1/2$ fermions as well as the presence of the hidden long-range ``string'' order despite the apparent  lack of distant spin-spin correlations~\cite{Nijs1989}. 
Despite numerous theoretical proposals~\cite{PhysRevLett.120.085301}, so far the Haldane phase has been only probed  by the neutron scattering \cite{Renard1991} or heat conductivity measurements \cite{Sologubenko2008,Kawamata2007} in anisotropic magnetic materials. 

Thus, the Rydberg excitons  provide a highly desirable table-top solid state platform with direct optical access to individual excitons for both fundamental studies 
 of hidden symmetries and topological orders and prospective quantum simulations~\cite{Demler2003,PhysRevB.85.075125,PhysRevLett.120.085301,RevModPhys.80.1083,RevModPhys.88.035005}.

\begin{figure}[b]
\includegraphics[width=0.45\textwidth]{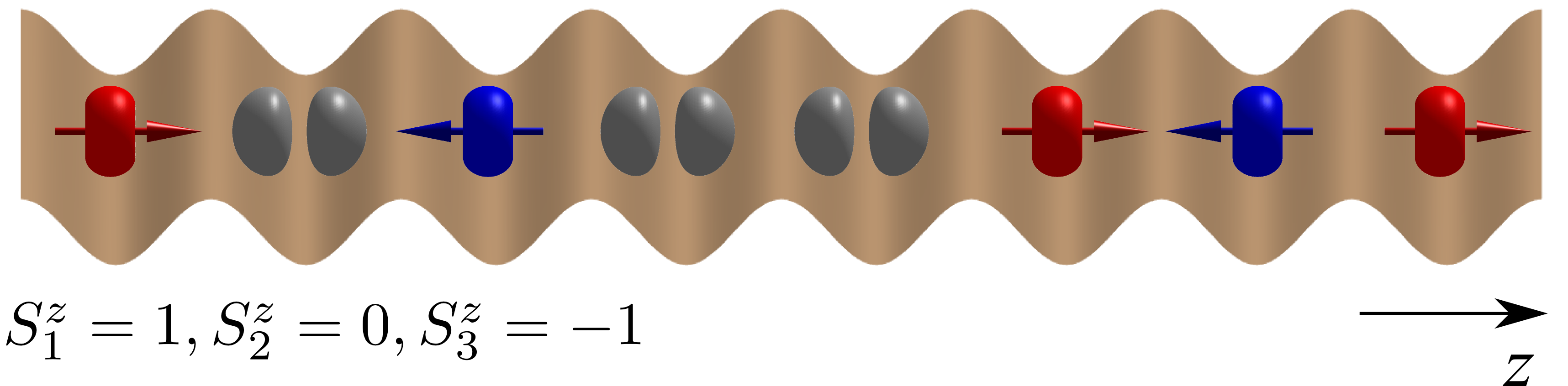}
\caption{  Schematics of the  diluted antiferromagnetic  state formed from $p$-shell Rydberg excitons in an array of traps}\label{fig:1}
\end{figure}

{\it {Chain} of Rydberg excitons.} We consider Rydberg excitons confined  in a one-dimensional {(1D)} periodic array  of traps, see Fig.~\ref{fig:1}.
{The traps for excitons could be created either optically utilizing the \emph{ac} Stark shift of excitons in the structured light waves, similar to the optical lattices of cold atoms~\cite{Bloch:2005aa} or in tailored  semiconductor environment where  the  band gap energy can be controlled  by applying the strain or external electrostatic potential~\cite{PhysRevB.34.2561}. It is assumed that each trap is occupied by a single exciton, that is in a $p$-shell state with the angular momentum  of the envelope function being equal to $1$. For simplicity, we neglect here exciton {internal} spin degrees of freedom, i.e., the spins of electrons and holes, and the spin-orbit interaction. 
We introduce the pseudovector angular momentum operator $\bm S_j=(S_j^x,S_j^y,S_j^z)$ for $j$th trap. Hereafter, we use the term spin to denote $\bm S_j$ for brevity, $z$ is the chain axis. We assume that the traps are sufficiently well separated, thus the excitons in neighboring traps are coupled to  each other {by} the van der Waals  interaction only. Owing to strong, $R^{-6}$, decay of the potential with the distance between the traps $R$, it is sufficient to consider the  nearest neighbor approximation. The Hamiltonian of the chain has the form ${\mathcal H}=\sum_{j=1}^{N-1} {\mathcal H}_{\rm bond}(\bm S_{j},\bm S_{j+1})\:,$
where $N$ is the number of  traps equal to the number of excitons in the system, $\mathcal H_{\rm bond}$ is the neighbors coupling Hamiltonian. The system has a rotational symmetry {around} the $z$ axis. 
{Thus, $\mathcal H_{\rm bond}$}  is characterized by $7$ real constants \cite{Klumper1993} {$c_0, \ldots, c_6$} and can be presented as
\begin{align}
\label{Hbond} 
{\mathcal H}_{\rm bond}(\bm S_{1},\bm S_{2})&=c_{0}+c_{1}S_{1}^{z}S_{2}^{z}+c_{2}
(S^{x}_{1} S^{x}_{2}+S^{y}_{1} S^{y}_{2})\\\nonumber &+c_{3}(S_{1}^{z} S_{2}^{z})^{2}+c_{4}
(S^{x}_{1} S^{x}_{2}+S^{y}_{1} S^{y}_{2})^{2}\\\nonumber
&+c_{5}[S_{1}^{z} S_{2}^{z} (S_{1}^{x} S_{2}^{x}+S_{1}^{y} S_{2}^y)+{\rm H.c.}]\\\nonumber&+c_{6}
(S_{1}^{x}S_{2}^{y}-S_{1}^{y}S_{2}^{x})^{2}\:.
\end{align}
The {microscopic calculation in Ref.~\cite{Walther2018} confirms that all $7$ parameters are significant.}
Specifically,
for the exciton {with}  principal quantum numbers $n={12\ldots 25}$ {one has}~\cite{Walther2018}
\begin{multline}
c_{0}=-5.58\mathcal E, c_{1}=9.53 \mathcal E, c_{2}=-8.97 \mathcal E, \\c_{3}=1.27\mathcal E, c_{4}=6.59\mathcal E, c_{5}=-3.18\mathcal E, c_{6}=5.04\mathcal E\:.\label{eq:C}
\end{multline}
where the common  factor $\mathcal E$ is $n^{15} \times 10/(2\pi)~ \rm mHz\times \mu m^{6}/R^{6}$.
{It is assumed that the trap size $d$ exceeds by far the exciton size but is much smaller than the distance between the traps $R$. In this situation the interaction Hamiltonian is independent of the trap size and geometry in the leading order in $d/R\ll 1$.} The Hamiltonian is invariant to the change of parameters $c_{2}\to -c_{2},c_{5}\to -c_{5}$ which corresponds to reflection $z\to -z$ for every second spin. 
Importantly, the spin-spin coupling is mostly antiferromagnetic ($c_{1}>0$) and strongly anisotropic.

Since the Rydberg blockade prevents two particles from being close to each other~\cite{gallagher2005rydberg,Saffman2010,Kazimierczuk2014}, in what follows we neglect the exciton tunneling between the traps. The interplay of the interactions and tunneling can enrich the spin phases in bosonic systems~\cite{PhysRevB.96.094435}.  The trap, however, can effect angular momentum state of the exciton giving rise to the anisotropic single particle contributions~\cite{supp}. We first disregard the anisotropy and discuss its effect in the end of the paper.

{\it Ground and excited state energies.}
The {chain} Hamiltonian can not be  diagonalized analytically for $N>3$. Instead, we first solved numerically for the lowest energy states in small finite chains  ($N\le 14$) with the periodic boundary {conditions~\footnote{We considered only chains with even $N$, since the odd number is incompatible with antiferromagnetic order, resulting in slower numerical convergence.}.}
\begin{figure}[t]
\includegraphics[width=0.45\textwidth]{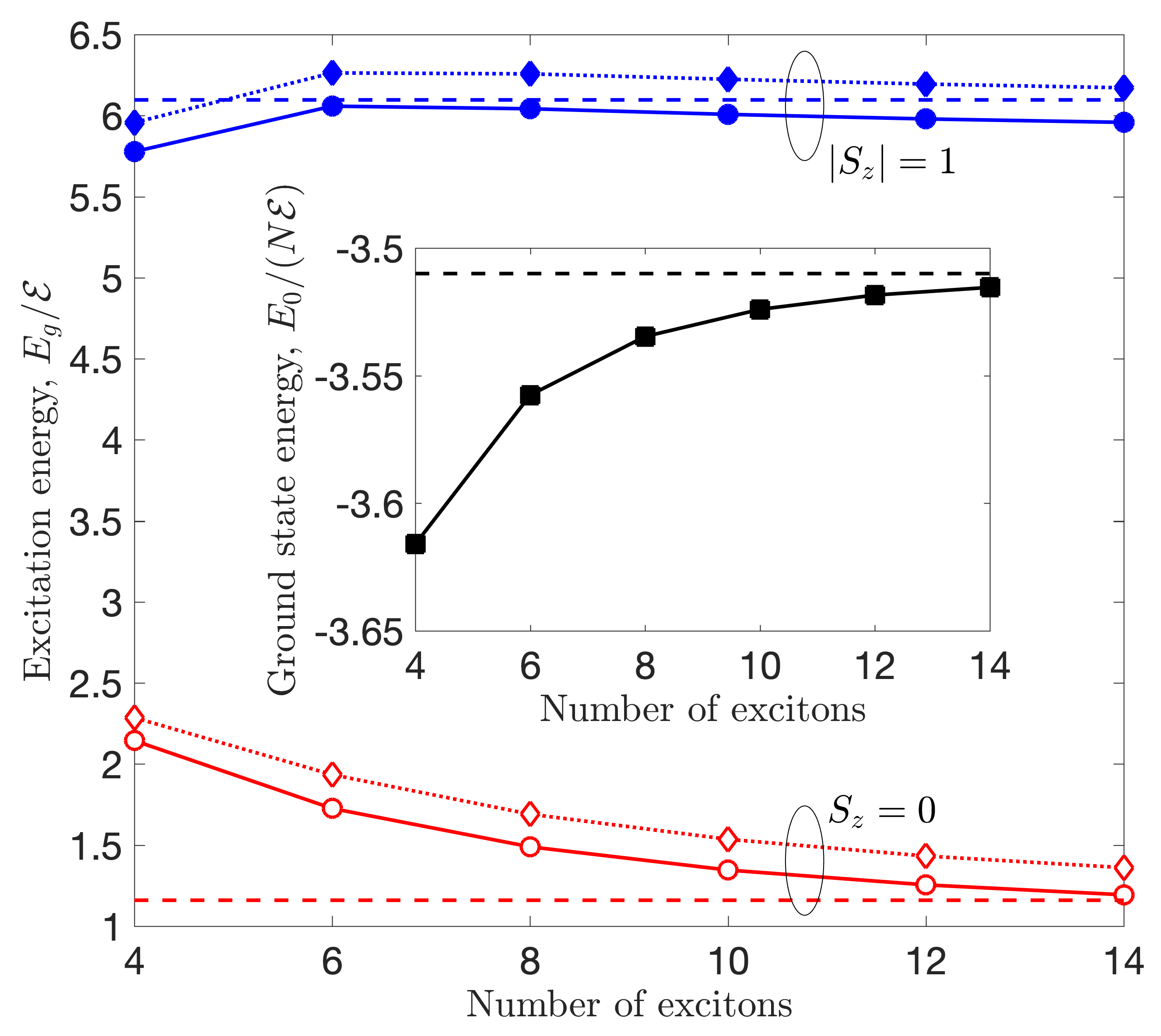}
\caption{  Ground and excited state energies in spin chains of different lengths with the periodic boundary conditions.
Open symbols correspond to the lowest excited state with the total spin projection $\sum S_{j}^{z}=0$, filled symbols correspond to the double-degenerate states with $\sum S_{j}^{z}=\pm 1$. Squares indicate results of direct numerical calculation, diamonds  correspond to the energies of the variational states Eqs.~\eqref{eq:SMA} and~\eqref{eq:SMA2} with $k=\pi$ {and $0$, respectively}. Dashed lines are the variational results obtained for the infinite system. Inset shows the ground state energy per bond, $E_0/(N\mathcal E)$, for the finite chain (symbols) and the infinite chain 
(dashed line).
}\label{fig:2}
\end{figure}
The results of calculation are shown in Fig.~\ref{fig:2}. The ground state energy, shown in the inset, quickly converges to the value
of $E_{0}\approx -3.51 N\mathcal E$.
The ground state is non-degenerate for periodic boundary conditions and has the total spin  $\sum_{j} S_{j}^{z}=0$ . The system is gapped, with the excitation gap slowly decreasing with the  size. The lowest excited state has zero total spin projection and the excitation energy $E_{g}\gtrsim 1.1\cal E$ (red open symbols in Fig.~\ref{fig:2}). The states with $\sum_{j} S_{j}^{z}=\pm 1$ have significantly  larger excitation energies ${E_{g}'}\gtrsim 6\cal E$ (blue filled symbols). In order to reveal the structure of the excited states we have compared the  results of exact diagonalization with those obtained from  the single-magnon variational {approach}~\cite{Arovas1988,Bartel2003}. Namely, we 
used the trial wavefunction 
 \begin{equation}
\psi_{k}\propto \frac{1}{\sqrt{N}}\sum\limits_{j=1}^{N}\e^{\rmi kj}S_{j}^{z}\psi_{0} \label{eq:SMA}\:,
\end{equation}
for the first excited state, where $\psi_{0}$ is the numerically calculated ground state, and the wave vector has been set to $k=\pi$~\cite{supp}.
The variational energy is close to the exact one, which indicates that the  excited state is well described by the ansatz Eq.~\eqref{eq:SMA}. The large overlap {(about $0.99$)} between  Eq.~\eqref{eq:SMA} and the lowest excited state has been also confirmed numerically. 
The lowest excited states with the $\pm 1$ total spin projection {being degenerate due to the time-reversal symmetry were} sought  in the form
 \begin{equation}
\psi_{k}\propto \frac{1}{\sqrt{N}}\sum\limits_{j=1}^{N}\e^{\rmi kj}[\mu_{1}S_{j}^{z}+\mu_{2}S_{j}^{z}(S_{j}^{x}{\pm}\rmi S_{j}^{y})]\psi_{0} \:,\label{eq:SMA2}
\end{equation}
with $k=0$~\cite{supp} and $\mu_{1,2}$ being the variational coefficients. The energy of this trial state is close to the exact numerical result as well, cf. blue circles and diamonds in Fig.~\ref{fig:2}.

In addition to the exact diagonalization of small finite chains  we have also  employed the  infinite time-evolving block decimation (iTEBD) approach ~\cite{Vidal2003,Vidal2004,Vidal2007}. The advantage of this  numerical technique is that it  is capable to directly address infinite {chains}. It is based on the representation of the ground state {$\psi_0$} in the  matrix-product form \cite{Orus2014}
\begin{equation} 
\psi_{i_{1}i_{2}i_{3}\ldots } \propto M^{i_{1}}_{\alpha_{1}\alpha_{2}}M^{i_{2}}_{\alpha_{2}\alpha_{3}}M^{i_{3}}_{\alpha_{3}\alpha_{4}}\ldots\:,\label{eq:MPS}
\end{equation}
where $M$ is a certain 3-rd rank tensor, indices $i_{j}=0,+1,-1$ label the projections of the spin $j$ and $\alpha_{j}=1\ldots \chi$ are auxiliary indices. If the spins were independent, one could use $\chi=1$, so that the $\alpha$ indices are suppressed, $M_{\alpha_{j}\alpha_{j+1}}^{i}\to M^{i}$ and the state Eq.~\eqref{eq:MPS} would reduce to a simple product state. For $\chi>1$ the state Eq.~\eqref{eq:MPS} can describe quantum entanglement {(i.e., correlations)} between the spins.
The iTEBD approach enables high-accuracy  description of arbitrary   gapped noncritical 1D systems  with local Hamiltonians \cite{Vidal2003b,Orus2014} {at relatively} low computational cost. In our case the convergence of energy better than 1\% has been already reached for the rank $\chi\lesssim 20$.

The iTEBD results for the infinite system are shown in Fig.~\ref{fig:2} by dashed horizontal lines. The energies of both ground and excited states well agree with the corresponding results for the finite chains. The excited state energies in the infinite system were estimated using the same variational  approximation Eqs.~\eqref{eq:SMA} {and} \eqref{eq:SMA2} with $\psi_{0}$ being now the ground state found from iTEBD.  We have also calculated the dispersion of excitations  $E_{g}(k)$ in the  single magnon approximation~\cite{supp}. In agreement with the analysis of the finite system, the spin-0 magnon branch is {indeed} gapped with the minimum at the edge of the Brillouin zone, $k=\pi$, while the magnons with $\pm 1$ spin projection have the minimum energy at $k=0$.

\begin{figure}[t]
\includegraphics[width=0.45\textwidth]{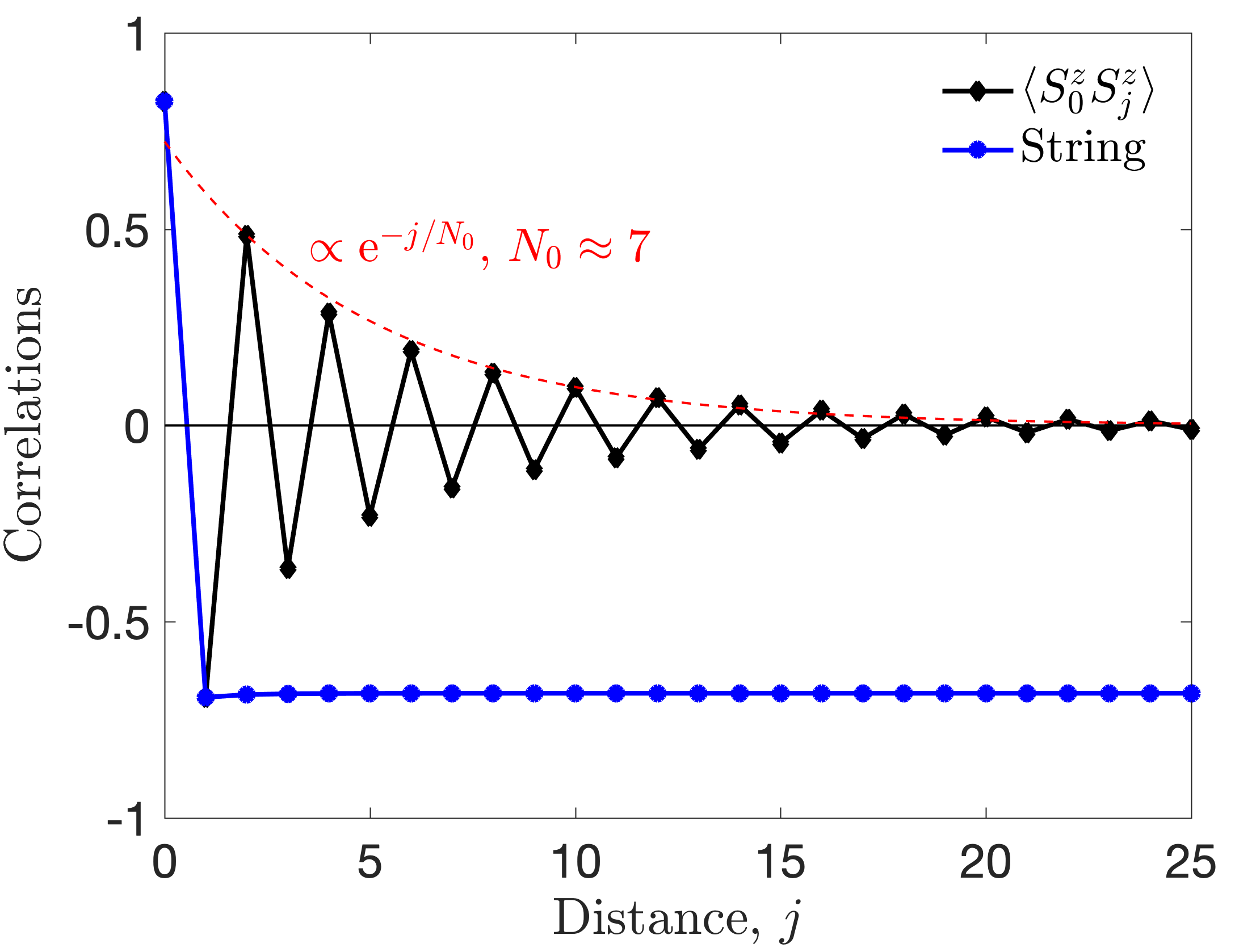}
\caption{Spin-spin correlator $(-1)^{j}C_{\text{N\'eel}}(j)$ and string correlator  for ground state of the infinite chain of 
Rydberg excitons, Eqs.~\eqref{eq:Neel} and~\eqref{eq:string}. 
}\label{fig:3}
\end{figure}

\begin{figure}[t]
\includegraphics[width=0.45\textwidth]{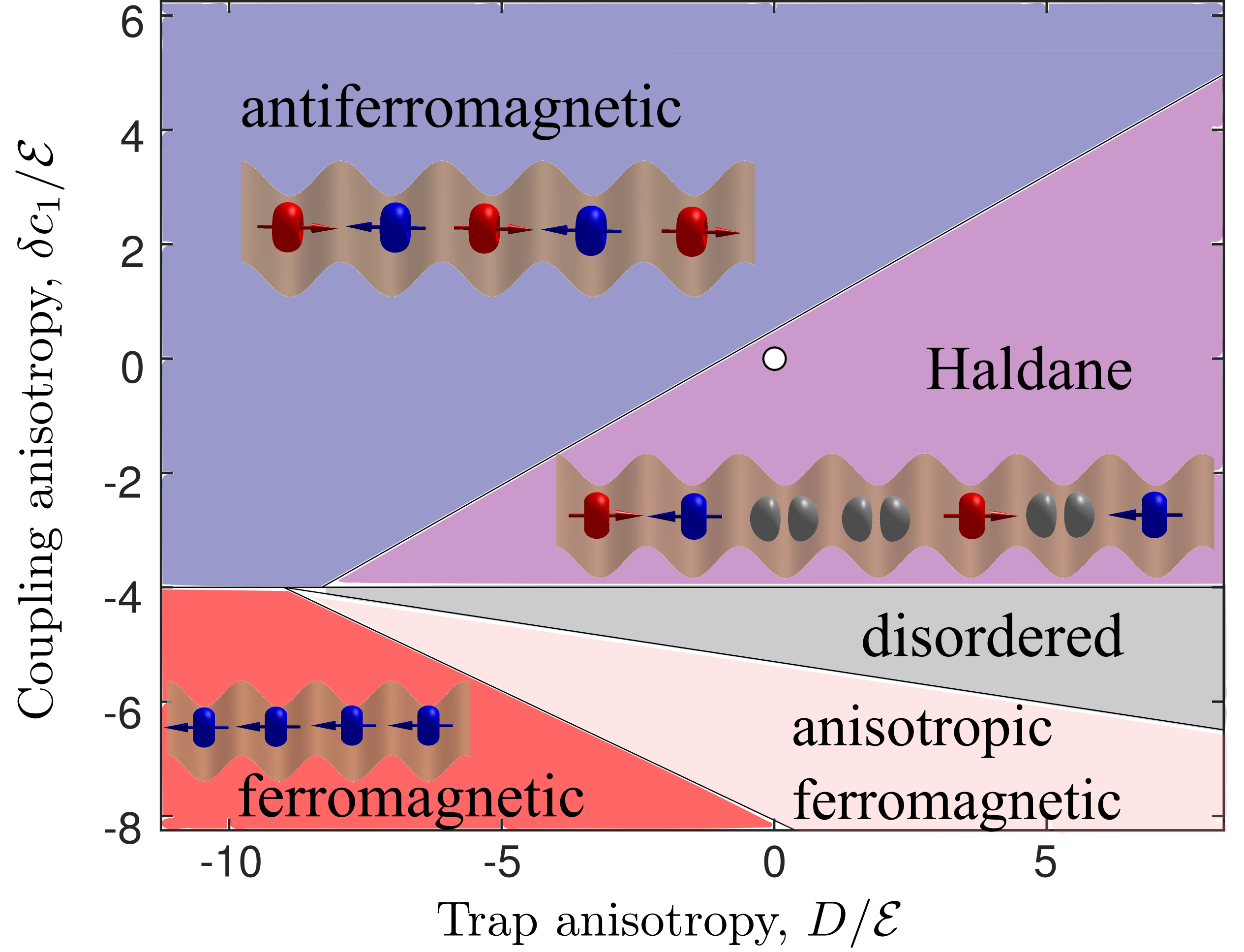}
\caption{Phase diagram  of the ground state depending on the
trap anisotropy and the spin-spin coupling anisotropy. Insets illustrate the typical spin configurations. {White circle indicates the parameters of Ref.~\cite{Walther2018} ($D=\delta c_1 =0$).}}\label{fig:4}
\end{figure}
{\it Spin structure of the ground state.} Now we proceed to the analysis of the spin-spin correlations in the ground state. The
calculation has been performed for the infinite chain using the iTEBD technique. The results are shown in Fig.~\ref{fig:3}. 
We start with the analysis of the pair spin-spin {correlations
depending} on the spin-spin distance $j$. The calculation demonstrates the presence of short-range antiferromagnetic N\'eel order, 
\begin{equation}
{C_{\rm{N\acute{e}el}}(j)\equiv (-1)^{j}\langle S_{0}^{z}S^{z}_{j}\rangle}\:,\label{eq:Neel}
\end{equation}
At large distances the  spin correlations Eq.~\eqref{eq:Neel} vanish, as illustrated by  the red dashed curve in Fig.~\ref{fig:3} showing the exponential decay with the correlation length $\approx 7$. Hence, based only on the analysis of local spin-spin correlations one could conclude that the considered spin phase has no long-range order.  However, this is not the case. Our main result is that the infinite chains of Rydberg excitons do possess a long-range string order, 
characterized by the non-vanishing correlator~\cite{Nijs1989}
\begin{equation}
C_{\rm string}(j)=\langle S_{0}^{z}\e^{\rmi \pi (S_{1}^{z}+S_{2}^{z}+\ldots S_{j-1}^{z})}S_{j}^z\rangle\:,\label{eq:string}
\end{equation}
blue circles in Fig.~\ref{fig:3}. The presence of the  non-local order parameter $C_{\rm string}(j)$ can be interpreted as a result of spontaneous breaking of a certain $\mathbb Z_{2}\times \mathbb Z_{2}$ symmetry, hidden in the system~\cite{Kennedy1992}.
Such string order in the absence of N\:eel order has been first revealed in the Afflect-Kennedy-Lieb-Tasaki (AKLT) model with ${\mathcal H_{\rm bond}}=\bm S_{1}\cdot\bm S_{2}+(\bm S_{1}\cdot\bm S_{2})^{2}/3$ \cite{affleck1988,Kennedy1992b}.
The AKLT model is exactly solvable by the matrix product state Eq.~\eqref{eq:MPS} with {certain rank-2 matrices~\cite{Kennedy1992}. Similar anzatz}
has allowed us to obtain analytically the ground state energy with $E_{0}=-3.49 N\mathcal E$~\cite{supp}, which is very close to the numerically obtained value of $E_{0}=-3.51 N\mathcal E$ shown in the inset of Fig.~\ref{fig:2}.
The simultaneous  presence of the string order and the absence of the N\'eel order, along with the presence of the non-degenerate gapped ground state  are the clear  fingerprints of the Haldane phase of spin-1 excitons~\cite{Haldane1983,Haldane1983a}. Physically, this indicates so-called diluted antiferromagnetism where the wavefunction of the ground state $\psi_0$ can be represented as a superposition of basic functions in the form
\[
|\ldots,0\ldots, 0, 1, -1, 0, \ldots, 1,-1,1,0 \ldots\rangle,
\]
where the significant part of {excitons} is in the $S^z=0$ state, and the spins $S^{z}=\pm 1$ can occur only in pairs, i.e., $+1,-1$ or $-1,+1$, while the pairs $+1,+1$ or $-1,-1$ are forbidden. Thus, the string correlator singles out ``diluted {anti}ferromagnetic order'', where neighboring spins are always opposite with the possible arbitrary number of zeroes in between. The fact that the pair $+1,-1$ or  $-1,+1$ can occur at any arbitrary place of the chain is related to the formation of the topological Haldane phase in our system.
Such diluted antiferromagnetic state is schematically illustrated in Fig.~\ref{fig:1}.

The shooting gun evidence for the Haldane phase is the presence of  edge states in a finite chain with open boundary conditions behaving as spin-$1/2$ fermions~\cite{Nijs1989,PhysRevB.85.075125}. Our numerical calculations for $N\leqslant 14$ exciton chains have indeed confirmed that, in the case of open boundary conditions, the ground state corresponds to $4$ closely lying levels corresponding to the combinations of edge spins-$1/2$ slightly split due to the finite chain size~\cite{supp}. 

\emph{Phase diagram}. The Haldane phase is a robust generic feature of 1D spin-1 chains with antiferromagnetic nearest-neighbor interaction. For instance, it is known to exist in isotropic  bilinear-biquadratic spin chains with ${\mathcal H}_{\rm bond}=-\cos\theta\bm S_{1}\cdot\bm S_{2}+\sin\theta (\bm S_{1}\cdot\bm S_{2})^{2}$ in a wide range of angles around $\theta=\pi$, including the spin-1 anisotropic  Heisenberg model~\cite{schollwock2008quantum}. In order to demonstrate, that the formation of the Haldane phase for the Rydberg excitons is not a  coincidence, we have analyzed the structure of the ground state depending on the anisotropy of the trap shape and on the sign of the coupling, either ferromagnetic or antiferromagnetic. The trap anisotropy has been described by adding the {single particle} terms $\mathcal H_1 = D\sum_j [(S_j^{z})^2-2/3]$ to the Hamiltonian~\cite{supp}. The coupling anisotropy was described by additional term $\delta c_{1}S_{1}^{z}S_{2}^{z}$ in the bond Hamiltonian Eq.~\eqref{Hbond}. The structure of the ground state was determined from comparison of the  spin-spin and string correlations at large {distances}. The results of this analysis are summarized in Fig.~\ref{fig:4}, and the correlation functions are given in  the Supplementary Materials~\cite{supp}.

We start the discussion of the phase diagram Fig.~\ref{fig:4} with the role of the coupling  term $\delta c_{1}S_{1}^{z}S_{2}^{z}$. Clearly, for large positive $\delta c_{1}$ the spin-spin interaction becomes antiferromagnetic, $\langle S_{1}^{z}S_{2}^{z}\rangle=-1 $, while for negative $\delta c_{1}$ the system is driven in the ferromagnetic phase with $\langle S_{1}^{z}S_{2}^{z}\rangle=1 $. 
The anisotropic ferromagnetic order, indicated by pink area in Fig.~\ref{fig:4}, is characterized by the long-range spin-spin correlations with
$\langle S_{1}^{z}S_{2}^{z}\rangle<1$.  The Haldane phase is realized  in the antiferromagnetic regime 
in the wide range of the  trap anisotropy parameter $D$ provided that it is  not large negative. {The impact of the anisotropy has also a transparent interpretation: Large positive values of $D$ favor $S^z=0$ ground state first facilitating the formation of the diluted antiferromagnetic state and, ultimately, the non-magnetic state with $S^z_j \equiv 0$ at each site. Similarly to the spin-$1$ Heisenberg model even slight anisotropy  $D<0$ switches the system into the antiferromagnetic state~\cite{Hida2003}. Large negative values of $D$ push down $S^z=\pm 1$ states rendering the exciton chain to the set of $1/2$-pseudospins and suppressing the Haldane phase.

%



\emph{Acknowledgements.} We are grateful to M. A\ss mann, M. Bayer, T. Pohl, M.A. Semina, and V. Walther for fruitful discussions. 

\nocite{apsrev41Control}
\bibliographystyle{apsrev4}

\end{document}


\title{Supplementary Materials for \\``Topological spin phases of trapped Rydberg excitons in Cu$_2$O'' }
\author{A.N. Poddubny and M.M. Glazov}
\affiliation{Ioffe Institute, St. Petersburg 194021, Russia}
\date{\today}
\maketitle 
\tableofcontents

\section{Variational approximation for the ground state energy}\label{sec:var}
\begin{figure}[h]
\includegraphics[width=0.4\textwidth]{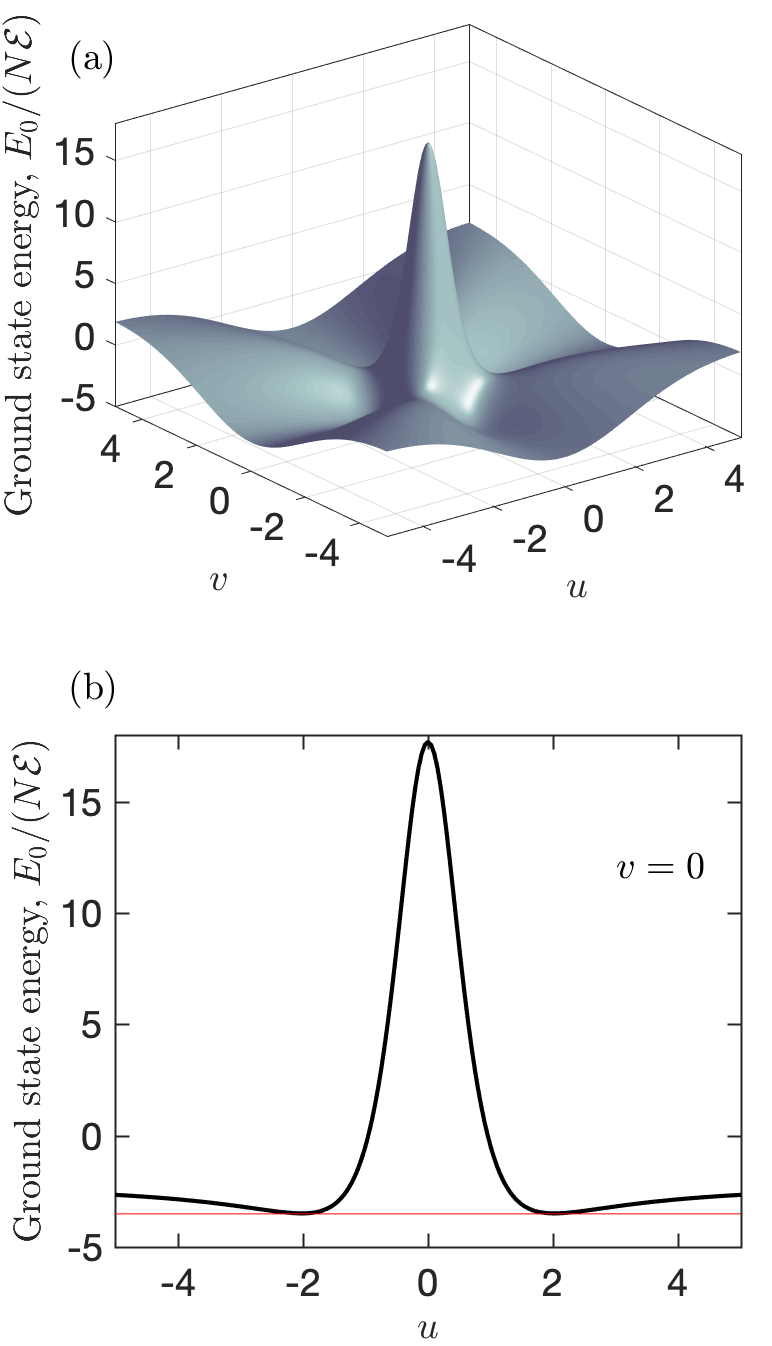}
\caption{Ground state energy depending on the variational  parameters $u$ and $v$ in the state~\eqref{eq:ansatz},~\eqref{eq:ansatz2}. Panel (a) shows the dependence on both $u$ and $v$ {demonstrating a maximum at $u=v=0$ and four distinct minima}, panel (b) shows the dependence on $u$ {at} $v=0$. Horizontal red line in the panel (b) indicates the numerically calculated ground state energy $E_{0}=-3.51 N\mathcal E$ in the infinite chain. The calculation has been performed for the {same set of interaction constants as in the main text}.}
\label{fig:S1}
\end{figure}
In this section we {present} the variational  ansatz from Ref.~\cite{Kennedy1992b} that has allowed us to calculate  the ground state energy in the infinite chain with the precision of better than $1\%$. This approach is based on the application of a nonlocal transformation {(known as Kennedy-Tasaki transformation)}
\begin{equation}
{V=\prod_{j<k}\exp(\rmi \pi  S_{j}^{z}S^{x}_{k}) \label{eq:KT}}
\end{equation}
to the Hamiltonian of the chain. The key advantage of the transformation Eq.~\eqref{eq:KT} is that it effectively ``disentangles'' the spins. This  means that the ground state of the transformed  Hamiltonian $\widetilde{\mathcal H}=V \mathcal H V^{-1}$ can be sought  in a simple factorizable form.  {Following Ref.~\cite{Kennedy1992b} i}t is sufficient to consider{, after performing the transform in Eq.~\eqref{eq:KT}} a chain with just two sites and the periodic boundary conditions, and look for the ground state in the form
\begin{equation}\label{eq:ansatz}
\tilde\psi_{i_{1}i_{2}}=\phi_{i_{1}}\phi_{i_{2}}\:,
\end{equation}
where $i_{1}, {i_2}=1,0,-1$ are the spin {components} and the {``trial''} spinor $\phi$ reads
\begin{equation}\label{eq:ansatz2}
\phi=\frac{1}{\sqrt{u^{2}+v^{2}+1}}\begin{pmatrix} u\\0\\v\end{pmatrix}\:.
\end{equation}
The ground state energy is  obtained by minimizing the expectation value of $\langle {\tilde\psi}|\widetilde {\mathcal H}|{\tilde\psi}\rangle$ as a function of $u$ and $v$. The dependence of this  expectation value on the variational parameters is shown in Fig.~\ref{fig:S1}. There exist four distinct minima for $u=\pm 2,v=0$, $v=\pm 2, u=0$. The
presence of four minima for $u,v\ne 0$ can be interpreted as a spontaneous symmetry  breaking effect. This so-called  hidden $\mathbb Z_{2}\times \mathbb Z_{2}$ symmetry is not obvious in the original Hamiltonian $
{\mathcal H}$ but is uncovered by the nonlocal Kennedy-Tasaki transformation. The minimum energy is 
$E_{0}=-3.49 N\mathcal E$, which is  close to the exact value for the infinite chain $E_{0}=-3.51 N\mathcal E$ calculated by the iTEBD  technique, cf. the black curve  and the red line in Fig.~\ref{fig:S1}(b).

Applying the operator Eq.~\eqref{eq:KT} to the state Eq.~\eqref{eq:ansatz} with $v=0$, we obtain  a ground state $V\tilde \psi$ in a matrix-product form
\begin{equation} 
\psi_{i_{1}i_{2}i_{3}\ldots } \propto M^{i_{1}}_{\alpha_{1}\alpha_{2}}M^{i_{2}}_{\alpha_{2}\alpha_{3}}M^{i_{3}}_{\alpha_{3}\alpha_{4}}\ldots\:,\label{eq:MPS}
\end{equation}
with
\begin{align}
M^{1}&=\begin{pmatrix}
0&0\\-\frac{u}{\sqrt{u^{2}+1}}& 0
\end{pmatrix},\quad M^{-1}&=\begin{pmatrix}
0&\frac{u}{\sqrt{u^{2}+1}}\\0& 0\end{pmatrix}\\
M^{0}&=\begin{pmatrix}
-\frac{1}{\sqrt{u^{2}+1}}&0\\0& \frac{1}{\sqrt{u^{2}+1}}
\end{pmatrix}\:, \nonumber
\end{align}
where the indices $i_{j}=0,+1,-1$ label the projections of the spin $j$.
The rank-two matrices $M$ describe the correlations and entanglement between the spins. {Thus, as expected from general analysis, the variational approach of Ref.~\cite{Kennedy1992b} gives a simplified version of the matrix-product wavefunction.}

\section{Dispersion of the excitations}\label{sec:magnon}
Here we present the {variational} calculation of the dispersion $E_{g}(k)$ of the lowest excitations in the infinite exciton chain. 
\begin{figure}[t]
\includegraphics[width=0.45\textwidth]{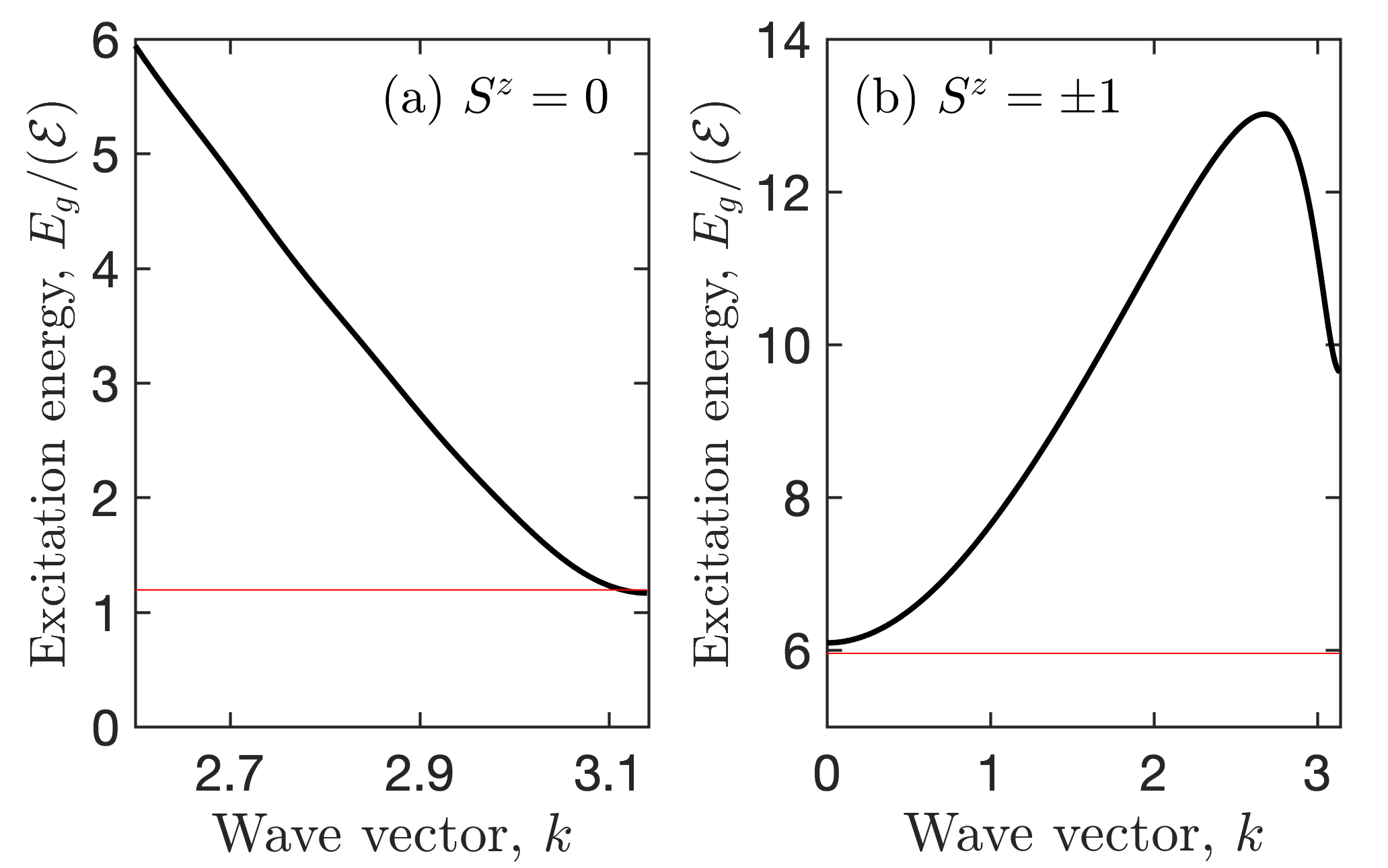}
\caption{Dispersion of the excitations with the total spin $0$ (a) and the total spin $1$ (b) in the infinite chain.
Horizontal red lines  show the excitation gap calculated numerically for a finite chain with $N=14$ excitons  $E_{g}=1.19\mathcal E$ {(a) and the gap to the $|S_z|=1$ excitations $E_g = 5.96\mathcal E$~(b)}. The calculation has been performed in the single-magnon approximation.}
\label{fig:S2}
\end{figure}

It has been proved that the  spectrally separated excitations  of a non-critical translation-invariant system can be  arbitrarily well approximated by building a momentum superposition of a local operator acting on the ground state~\cite{Haegeman2013}. For our purposes it is sufficient to use  the most simple single-magnon approximation~\cite{Arovas1988,Bartel2003}. The excitation with the total spin equal to zero is sought  in the following form 
 \begin{equation}
\psi_{k}\propto \frac{1}{\sqrt{N}}\sum\limits_{j=1}^{N}\e^{\rmi kj}S_{j}^{z}|0\rangle \label{eq:SMA}\:,
\end{equation}
where $|0\rangle$ is the numerically calculated ground state.
The excitation energy is then given by 
\begin{equation} \label{eq:SMA2}
E_{g}(k)=\frac{s(k)}{2f(k)} 
\end{equation}
where
\begin{equation}
s(k)=\frac{1}{N}\sum\limits_{jj'}\e^{\rmi k(j-j')}\langle 0|[S_{j'}^{z'},[H,S_{j}^{z}]]|0\rangle\:
\end{equation}
and 
\begin{equation}
f(k)\equiv \langle 0| S_{-k}^{z}S_{k}^{z}|0\rangle=\frac{1}{N}\sum\limits_{jj'}\e^{\rmi k(j-j')}\langle 0|S_{j'}^{z'}S_{j}^{z}|0\rangle
\end{equation}
is the structure factor. The main advantage of the single-magnon approximation is that it directly expresses the excitation energy via the spin-spin correlations in the ground state. Hence, once the ground state is obtained by the iTEBD technique, the dispersion of excitations can be calculated without {significant} additional numerical cost.
The precision can be further improved by expanding  the set of the variational functions~\cite{Haegeman2013b,Zauner2015}. {The results of variational calculations in the $S_z=0$ sector of excitations are shown in Fig.~\ref{fig:S2}(a) in the vicinity of $k=\pi$, where the minumum of the dispersion is realized.}

In the case of spin-1 excitation, the variational function  should be described by the two parameters~\cite{Bartel2003}
 \begin{equation}
\psi_{k}\propto \frac{1}{\sqrt{N}}\sum\limits_{j=1}^{N}\e^{\rmi kj}[\mu_{1}S_{j}^{z}+\mu_{2}S_{j}^{z}(S_{j}^{x}{\pm}\rmi S_{j}^{y})]\psi_{0} \:,\label{eq:SMA2}
\end{equation}
The excitation energy is then found from a generalized $2\times 2$ eigenproblem 
\begin{equation} \label{eq:SMA2}
\mathcal H(k) \psi= E_{g}(k) {\mathcal W} \psi,
\end{equation}
where 
\begin{align}
\mathcal H_{\alpha\beta}(k)=&
\sum\limits_{jj'}\e^{\rmi k(j-j')}\langle 0|O^{\alpha}_{j}HO_{j}^{\beta}|0\rangle,\\
\mathcal W_{\alpha\beta}(k)=&
\sum\limits_{jj'}\e^{\rmi k(j-j')}\langle 0|O_{j'}^{\alpha}O_{j}^{\beta}|0\rangle
\end{align}
and $O_{j}^{1}=S_{j}^{z}$, $O_{j}^{2}=S_{j}^{z}(S_{j}^{x}+\rmi S_{j}^{y})$.
The results of the calculation are presented in Fig.~\ref{fig:S2}(b). Here the minimum is realized in the vicinity of the $k=0$ point. The variational results for the excitation gaps are in  agreement with the rigorous numerical calculation for small finite chains shown in Fig.~2 of the main text{, see also horizontal  red lines in Fig.~\ref{fig:S2}}.

\section{Role of the boundary conditions}\label{sec:var}

\begin{figure}[t]
\includegraphics[width=0.45\textwidth]{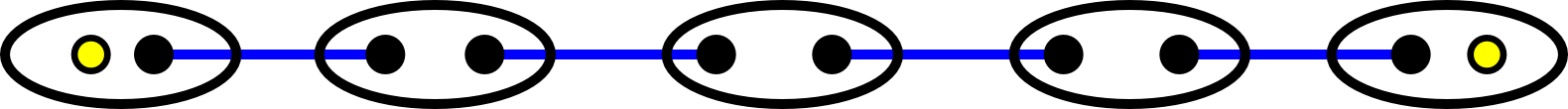}
\caption{Structure of  the ground eigenstate in the AKLT model with ${\mathcal H_{\rm bond}}=\bm S_{1}\cdot\bm S_{2}+(\bm S_{1}\cdot\bm S_{2})^{2}/3$~\cite{affleck1988,Kennedy1992b}. Each spin 1 (oval) is represented by triplet states of two spins 1/2 (dots). The spins 1/2 for neighboring sites are bound into singlets (blue lines). Two dangling spins remain in the finite chain (yellow dots) explaining the 4-fold degeneracy of the ground state in long chain with open ends.
}\label{fig:aklt}
\end{figure}
\begin{figure}[t]
\includegraphics[width=0.5\textwidth]{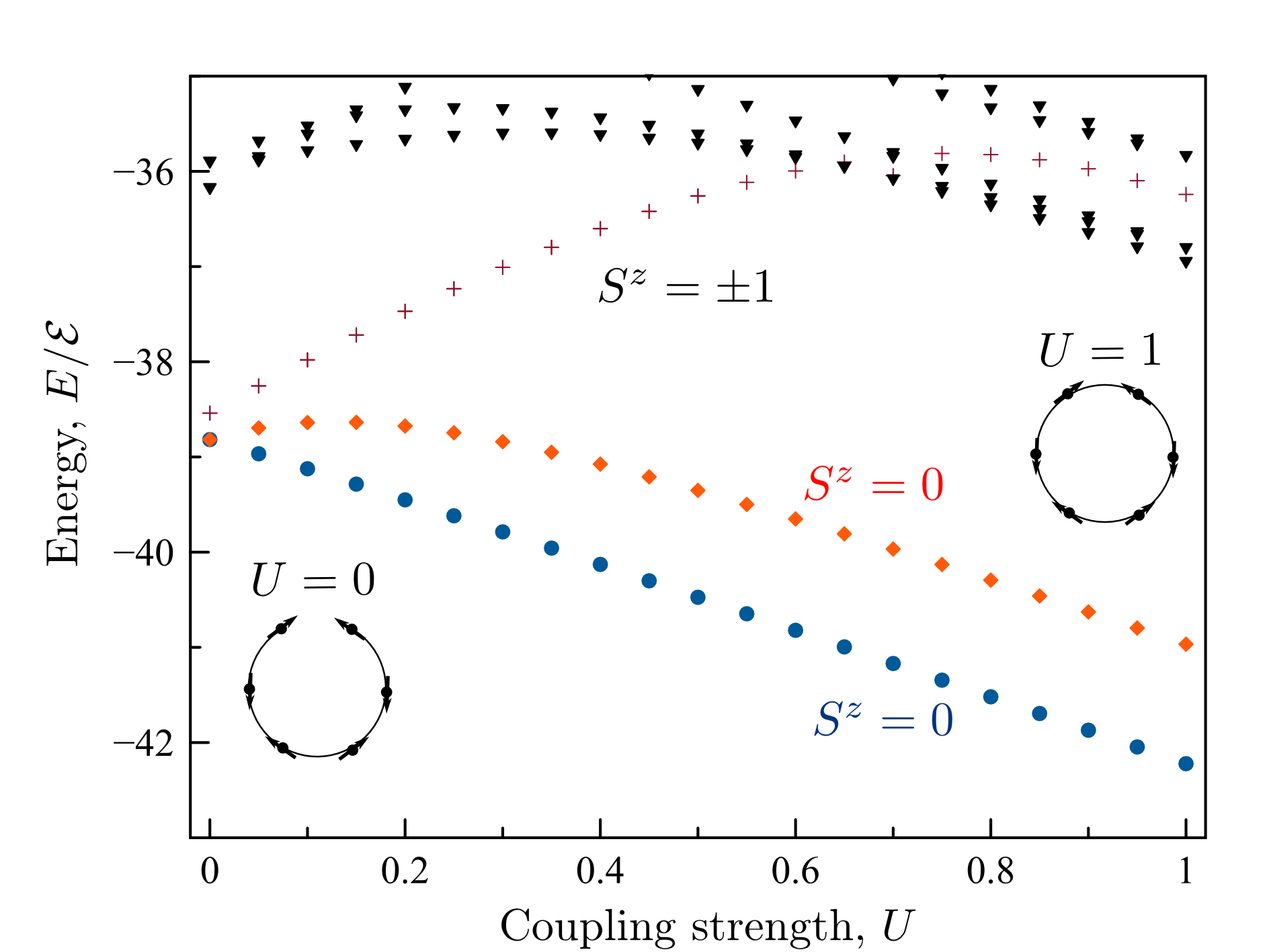}
\caption{Energies of the several lowest states in the exciton chain depending on the coupling parameter $U$ in Eq.~\eqref{eq:U}, controlling the boundary conditions.
The insets schematically illustrate the limiting cases of an open chain ($U=0$) and a ring ($U=1$).
Calculation has been performed for $N=12$ excitons.}
\label{fig:S3}
\end{figure}
The distinctive topological feature of the Haldane phase is the strong sensitivity of the ground state to the choice of boundary conditions, either open ones or periodic ones~\cite{affleck1988}. This effect can be understood in the simplest fashion for the isotropic Afflect-Kennedy-Lieb-Tasaki (AKLT) model with ${\mathcal H_{\rm bond}}=\bm S_{1}\cdot\bm S_{2}+(\bm S_{1}\cdot\bm S_{2})^{2}/3$.  Each of  the spins 1 (oval) can be equivalently replaced by {three-fold degenerate} triplet states of {\it two} {strongly interacting} spins 1/2 (circles). Next, the interaction {between the sites} leads to coupling of neighboring spins 1/2 (otherwise degenerate) into a singlet states. Two dangling spins (yellow dots) remain on the opposite edges of the finite chain, hence, the ground state becomes fourfold degenerate {in the infinite chain limit}, see Fig.~\ref{fig:aklt}.  When the open chain is closed and transformed to a ring, the dangling edge spins are fused together and the ground state becomes non-degenerate. {Similar effect is expected for the Haldane phase realized in our case of strongly anisotropic chain of Rydberg excitons.}

\begin{figure*}[th]
\includegraphics[width=0.85\textwidth]{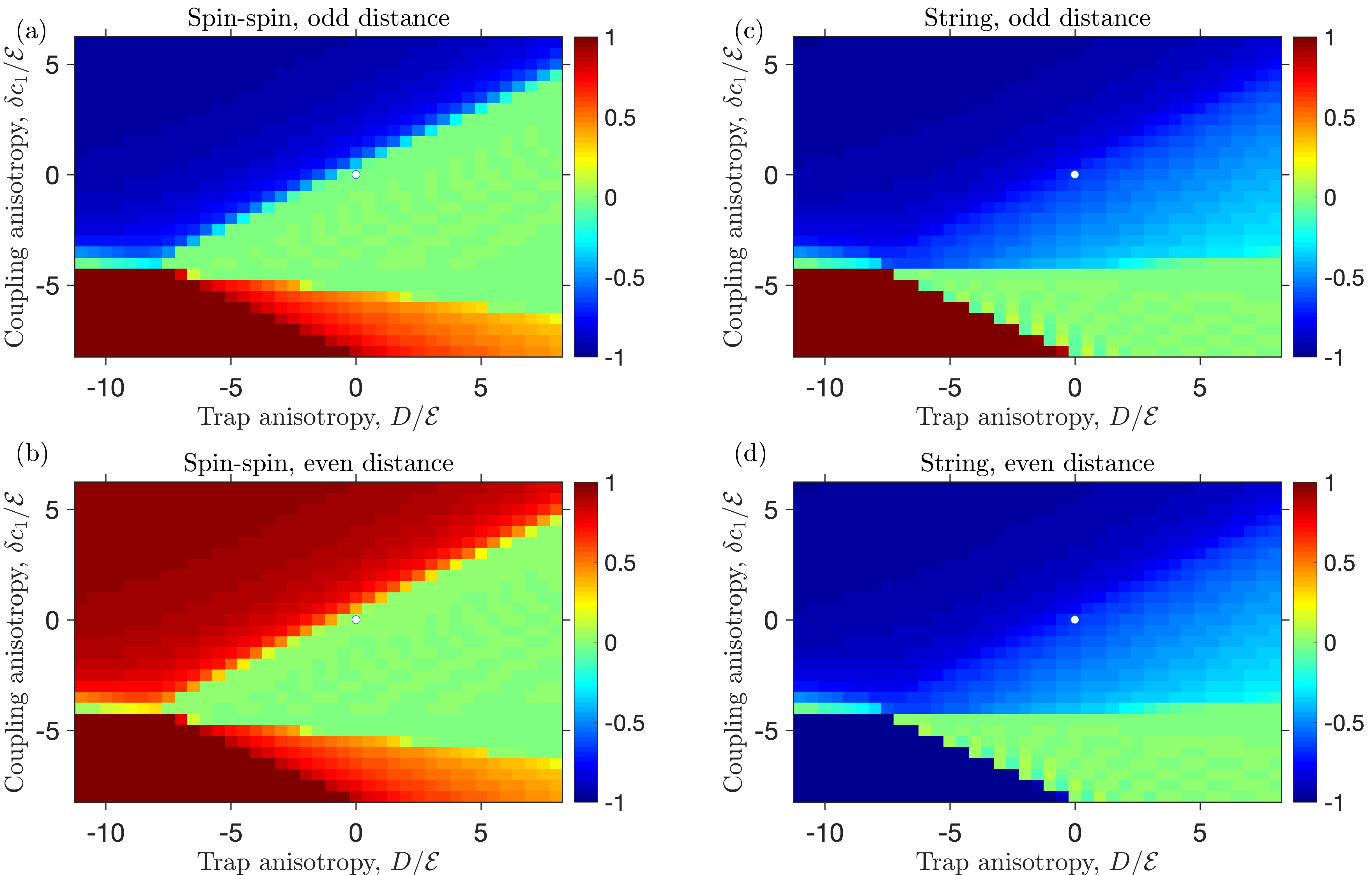}
\caption{Spin-spin correlators $\langle S_{0}^{z}S^{z}_{j}\rangle$ (a,c) and string correlators
$\langle S_{0}^{z}\e^{\rmi \pi (S_{1}^{z}+S_{2}^{z}+\ldots S_{j-1}^{z})}S_{j}^z\rangle$. Panels (a,b) and 
(c,d) correspond to odd and even spin-spin distances, $j=39$ and $j=38$, respectively. {White circle indicates the parameters of Ref.~\onlinecite{Walther2018} ($D=\delta c_1 =0$).}}
\label{fig:S4}
\end{figure*}

 In order to examine these dramatic boundary effects in our system, we have explicitly analyzed the transition between an open chain and and a closed ring, both described by a  Hamiltonian
\begin{equation}
{\mathcal H}=\sum_{j=1}^{N-1} {\mathcal H}_{\rm bond}(\bm S_{j},\bm S_{j+1})+U {\mathcal H}_{\rm bond}(\bm S_{N},\bm S_{1})\:. \label{eq:U}
\end{equation}
Here,  the parameter $U$ controls the boundary conditions: the value  of $U=1$ corresponds to the periodic boundary conditions, i.e. spins on a ring. When $U=0$, the ring is broken and the boundary conditions correspond to an open chain. Figure~\ref{fig:S3} presents the energy spectrum calculated by full diagonalization for a small chain with $N=12$ excitons. It demonstrates that the ground state of the open chain has almost exact 4-fold degeneracy. The four bottom levels can be understood as the combinations of edge spins-$1/2$.  Namely, four spins $1/2$ can form a singlet with a zero total spin and a triplet with the total spin $S$ equal to one. The triplet states with $S^{z}=\pm 1$ are degenerate due to the time inversion symmetry. The two states with $S^{z}=0$, stemming from a singlet and a triplet, are slightly split due to the finite size of the chain.  When the coupling parameter $U$ is increased, the ground state is dramatically modified and transforms from an (almost) four-fold degenerate one to a non-degenerate one, reflecting the  fusion of the edge $1/2$  spins. The energy of one of the states with $S^{z}=0$ decreases by $\approx E_{0}=-3.5\mathcal E$, since an  an extra bond is formed in a system. The second state with $S^{z}=0$, being degenerate with the ground state in the open chain, becomes in a first excited state of the close ring, corresponding to a  magnon mode from Fig.~\ref{fig:S2}(a). The states with $S^{z}=1$ increase in energy and are also transformed to the magnon modes from Fig.~\ref{fig:S2}(b). Importantly, the comparison of the limits $U=0$ (open chain) and $U=1$ (ring) shows that the degeneracy of the ground state is higher for the open chain being an evidence of edge uncoupled spins appearing in the system.

\section{Ground state phase diagram}\label{sec:phase}
In this section we present the {numerical} data behind the phase diagram of the ground state, Fig.~4 in the main text. In order to obtain the phase diagram we have calculated the spin-spin and string correlators in the ground state at large  odd and even distances {($j = 39$ and $j=38$, respectively) which exceed by far the correlation lengths in our system ($N_0\sim 7$, Fig. 3 of the main text). The results are presented in Fig.~\ref{fig:S4}. The Haldane phase corresponds to the vanishing spin-spin and nonzero string correlations, white circle in Fig.~\ref{fig:S4}. On the other hand, in the ferromagnetic and antiferromagnetic phases both {string correlations and spin-spin correlations are nonzero.  The phases} can be distinguished by the relative sign of the correlations for $j$ and $j+1$, positive for the ferromagnetic phase and negative for the antiferromagnetic phase. The disordered phase does not display correlations.

{\section{Rydberg exciton states in a trap}}\label{sec:trap}
The single particle states of the Rydberg exciton in a trap are described by the Hamiltonian
\begin{equation}
\mathcal H_x =- \frac{\hbar^2}{2\mu} \Delta_{\bm \rho} - \frac{e^2}{\varkappa\rho} - \frac{\hbar^2}{2M} \Delta_{\bm R} + V(\bm \rho, \bm R),\label{H:trap}
\end{equation}
acting on the wavefunction of the electron-hole pair, $\Psi(\bm \rho, \bm R)$. Here we use the center-of-mass, $\bm R = (m_e\bm r_e+m_h\bm r_h)/M$, and relative motion, $\bm \rho = \bm r_e - \bm r_h$, coordinates, $\mu$ and $M$ are the reduced and translational masses of the electron-hole pair, respectively, $m_e$ and $m_h$ are the masses of the electron and the hole, $\bm r_e$ and $\bm r_h$ are their coordinates, $\varkappa$ is the dielectric constant. In Eq.~\eqref{H:trap} $V(\bm \rho, \bm R) = V_e(\bm r_e)+ V_h(\bm r_h)$ is the confinement potential being the sum of the electron and hole potential energies in the trap. 

In the absence of the trap the exciton wavefunction is decomposed as a product of the center of mass and relative motion wavefunctions, $\Psi(\bm \rho, \bm R) = \psi(\bm R) \Phi(\bm r)$. The states of the relative motion
\begin{equation}
\label{relative}
\Phi_{n,l,m}(\bm \rho) = R_{nl}(\rho) Y_{lm}(\vartheta, \varphi),
\end{equation}
form hydrogen-like series with $n=1,2, \ldots$ being the principal quantum number, $l=0, 1, \ldots n$ being the orbital angular momentum and $m=-l,\ldots,l$ being its component or magnetic quantum number~\cite{ll3_eng,Kazimierczuk2014}; $\rho = |\bm \rho|$, $\vartheta$ and $\varphi$ are the angles of the $\bm \rho$. The effects related to the crystalline anisotropy are disregarded~\cite{Glazov2015,PhysRevB.93.195203}.

We take the trap potential in the simplest possible form assuming that electron and hole are confined by the parabolic potential.  The presence of the trap does not allow one, in general, to separate the electron and hole coordinates, however, for particular choice of the trap parameters (where the oscillator frequencies for both carriers are the same\cite{que92,Semina2006}) the trap potential can be recast as
\begin{multline}
\label{trap}
V(\bm \rho, \bm R) = \frac{\mu \Omega_\perp^2}{2}(x^2+y^2) + \frac{\mu \Omega_\parallel^2}{2}z^2\\
 + \frac{M \Omega_\perp^2}{2}(X^2+Y^2) + \frac{M \Omega_\parallel^2}{2}Z^2.
\end{multline}
Here $z$ is the trap axis, $x$, $y$ and $z$ are the Cartesian components of the relative motion  position-vector, $\bm \rho$, and $X$, $Y$ and $Z$ are the Cartesian components of the center of mass motion position-vector $\bm R$, $\Omega_\parallel$ and $\Omega_\perp$ are the corresponding frequencies for the confinement along the trap axis and perpendicular to the trap axis, respectively. The center of mass quantization is described by the series of harmonic oscillator states with equidistant levels. 

For $\Omega_\parallel = \Omega_\perp$ the relative motion states can be still represented in the form of Eq.~\eqref{relative} due to spherical symmetry of the problem. In this case three $p$-shell states, $p_x$, $p_y$ and $p_z$ corresponding to the linear combinations of $l=1$, $m=\pm 1$ and $0$ states are equivalent and have the same energies. By contrast, for $\Omega_\parallel \ne \Omega_\perp$ the splitting between the doublet $p_x$, $p_y$ and the singlet $p_z$ state appears. For the prolate trap, $\Omega_\parallel < \Omega_\perp$, the $p_z$ state has a lower energy, while for the oblate trap, $\Omega_\parallel > \Omega_\perp$, the $p_z$ state energy increases. The effective orbital angular momentum dependent Hamiltonian reads
\begin{equation}
\label{aniso:trap}
\mathcal H_1 = D \left(S_z^2 - \frac{2}{3}\right),
\end{equation}
where we excluded the overall energy shift of the $l=1$ triplet. The parameter $D $  is proportional to $\Omega_\parallel - \Omega_\perp$.

To quantify the scenario we assume that the trap potential is weak and can be treated by means of the first-order perturbation theory. Making use of the radial integrals we obtain 
\[
\int R_{nl=1}^2(\rho) \rho^4 d\rho = \frac{5}{2}n^2(n^2-1)\left(\frac{\hbar^2 \varkappa}{\mu e^2}\right)^2 ,
\]
we obtain for the parameter $D$ in Eq.~\eqref{aniso:trap}
\begin{equation}
\label{D}
D = \frac{\mu (\Omega_\parallel^2- \Omega_\perp^2)}{2} \left(\frac{\hbar^2 \varkappa}{\mu e^2}\right)^2 n^2(n^2-1).
\end{equation}

\nocite{apsrev41Control}
\bibliographystyle{apsrev4}